\documentclass[aps,prd,amsmath,twocolumn,amssymbaps,showpacs]{revtex4-2}
\usepackage{graphicx}  
\usepackage{dcolumn}   
\usepackage{bm}        
\usepackage{amssymb}   
\usepackage{amsmath}   
\usepackage{bbm}
\usepackage{draftcopy}
\usepackage{relsize} 
\usepackage{xfrac}    
\usepackage{slashed}
\usepackage{color}
\usepackage{footnote}
\definecolor{dgreen}{cmyk}{1.,0.,1.,0.2}        
\definecolor{orange}{cmyk}{0.,0.353,1.,0.}    

\usepackage[bookmarks]{hyperref}

\newcommand{\di}{{\rm d}}

\newcommand{\be}{\begin{equation}}
\newcommand{\ee}{\end{equation}}                                                                               
\newcommand{\bea}{\begin{eqnarray}}
\newcommand{\eea}{\end{eqnarray}} 

\begin{document}
\title{QCD phase transition at finite isospin density and magnetic field}

\author{Chujun Ke$^{1}$ and Gaoqing Cao$^{1,2}$}
\affiliation{1 School of Physics and Astronomy, Sun Yat-sen University, Zhuhai 519082, China\\
2 Guangdong Provincial Key Laboratory of Quantum Metrology and Sensing, Sun Yat-Sen University, Zhuhai 519082 
China}
\date{\today}

\begin{abstract}
The QCD phase transition is explored at finite isospin density and magnetic field within the extended two-flavor Nambu--Jona-Lasinio model. By adopting the Ginzburg-Landau approximation, we study the transitions from normal chiral symmetry breaking phase to pion superfluidity or rho superconductivity. To avoid the artificial divergence for a large isospin chemical potential, we adopt the Landau representation rather than the proper-time one for the fermion propagators in a constant magnetic field. For the Landau representation, the same cutoff to the Landau energies, rather than to Landau levels, should be adopted to regularize the divergences from the summations over Landau levels. Then, the Ginzburg-Landau coefficients for pion and rho mesons are worked out both analytically and numerically in random phase approximation. The results show that pion superfluidity is favored for a small magnetic field while rho superconductivity is favored for a large magnetic field when increasing isospin chemical potential, in line with the magnetic enhancement (reduction) of the lowest energy of $\pi^+ ({\rho}^{+})$ meson. The novel rho superconductivity phase at large magnetic field and finite isospin density implies an interesting and nontrivial interplay between QCD and QED.
\end{abstract}


\maketitle

\section{Introduction}

Quantum Chromodynamics (QCD) is the theory of strong interactions that governs the dynamics of quarks and gluons thus is fundamental to understand nuclear matter under extreme conditions \cite{Gross:1973id, Politzer:1973fx,Yagi:2005yb}. A central goal of modern nuclear physics is to utilize QCD or QCD-like effective theories to sketch rich phase diagrams of nuclear matter in the parameter space spanned by temperature, density, and others \cite{Klevansky:1992qe,Stephanov:2004wx, Fukushima:2010bq,Luo:2017faz,Son:2000xc,Kogut:2002zg,He:2005nk,Chen:2015hfc,Jiang:2016wvv,Cao:2021rwx}. These extremal conditions could usually be realized in compact stars~\cite{Lattimer:2004pg} and relativistic heavy-ion collisions \cite{Yagi:2005yb,Chen:2024aom,Chen:2026gka}; specially, strong magnetic fields, up to $10^{18}-10^{20} {\rm Gs}$, can be generated in magnetars~\cite{Bocquet:1995je} and non-central collisions \cite{Skokov:2009qp,Deng:2012pc}, see the experimental observations in Refs.~\cite{STAR:2023jdd, Huang:2024hua, Zhao:2022dac, Shen:2025unr}. Such strong magnetic fields could introduce profound anisotropy into the QCD system, quantize charged particles into Landau levels, and further modify the non-perturbative QCD vacuum. An important manifestation is the magnetic catalysis effect, that is, the magnetic field enhances chiral condensate at low temperatures \cite{Gusynin:1994xp}. Intriguingly, first-principles lattice QCD simulations reveal an inverse magnetic catalysis near the crossover temperatures, where the magnetic field suppresses the chiral condensate \cite{Bali:2011qj, Bali:2012zg}. On the other hand, since the spectra of charged hadrons are greatly affected by strong magnetic fields~\cite{Bali:2017ian,Ding:2020hxw,Ding:2026qzu}, instability might be developed in the traditional superfluid or superconductor phases, offering unprecedented chances to explore novel phases~\cite{Cao:2021rwx}.

At finite isospin chemical potential $\mu_{\rm I}$, it has been well established that the QCD vacuum would undergo a transition to a charged pion $(\pi^\pm)$ superfluidity for $\mu_{\rm I} > m_\pi$ \cite{Son:2000xc, Kogut:2002zg,He:2005nk,Brandt:2017oyy} and a BEC-BCS crossover at larger $\mu_{\rm I}$~\cite{Sun:2007fc,Cuteri:2021hiq}. While the isospin quantum numbers of charged rho mesons $\rho^\pm$ are the same as $\pi^\pm$, $\rho^\pm$ superconductivity is suppressed by $\pi^\pm$ superfluidity~\cite{Brauner:2016lkh}. However, due to different spins, they respond differently to a magnetic field: In the view of quasiparticles, the lowest energy of $\pi^\pm$ increases with $B$ while that of $\rho^\pm$ decreases with $B$~\cite{Cao:2020pmm}. The latter point led to the so-called vacuum superconductor of rho mesons \cite{Chernodub:2010qx, Chernodub:2011mc}, even though sequential studies show that such a phase would violate the Vafa-Witten theorem and there is no instability in the lowest energy of $\rho^\pm$ due to their internal quark components \cite{Hidaka:2012mz, Cao:2019res,Bali:2017ian,Ding:2020hxw}. Nevertheless, the difference implies that the lowest energy of $\pi^\pm$ could be larger than that of $\rho^\pm$ at some critical magnetic field, which has been verified in lattice QCD simulations~\cite{Bali:2017ian,Ding:2020hxw}, so magnetic field could help $\rho^\pm$ condensation to get an advantage over $\pi^\pm$ condensation. For instance, $\pi^\pm$ superfluidity was first predicted in parallel rotation and magnetic field~\cite{Liu:2017spl,Cao:2019ctl}, and then $\rho^\pm$ superconductivity was found to be more favored for a larger magnetic field~\cite{Cao:2020pmm}. By following the same logic, we would expect $\rho^\pm$ superconductivity to be more favored with increasing $\mu_{\rm I}$ at a larger magnetic field. For comparison, we note that rotation could polarize the spin of $\rho^\pm$ in a similar way as magnetic field, and it was discovered that rotation favors $\rho^\pm$ superconductivity at high $\mu_{\rm I}$ \cite{Zhang:2018ome, Zhang:2020drr}. 

If magnetic field favors $\rho^\pm$ superconductor, the later must be of type-II, otherwise Meissner effect would block any magnetic effect in type-I superconductor. So, the transition from normal phase to $\rho^\pm$ superconductivity should be of second order. It is not necessary the case for $\pi^\pm$ superfluidity, but the studies within chiral perturbation theory showed that $\pi^\pm$ superfluid is also of type-II thus the transition from normal phase is also of second order \cite{Adhikari:2015wva, Adhikari:2018fwm}. It is because that pions are so strongly bound that the penetration depth of magnetic field is much larger than the correlation length~ \cite{Adhikari:2015wva}. These observations justify the use of Ginzburg-Landau theory to study the phase transitions from normal phase to $\pi^\pm$ superfluidity or $\rho^\pm$ superconductivity in a magnetic field. In this sense, the phase boundary of $\pi^\pm$ superfluidity in $\mu_{\rm I}-B$ plane is valid in Ref.~\cite{Adhikari:2015wva} as long as $\rho^\pm$ is not involved. 
 
Motivated by these insights, this work aims to elucidate the intertwined effects of a constant magnetic field and finite isospin chemical potential on the QCD phase diagram, focusing specifically on the competition between $\pi^\pm$ superfluidity and $\rho^\pm$ superconductivity. This paper is organized as follows. In Sec.~\ref{NJL}, the whole formalism is developed: Formal derivations of gap equation and quadratic Ginzburg-Landau coefficients are given in Sec.~\ref{Formal}, and explicit evaluations and detailed regularizations are presented in Sec.~\ref{Explicit}. Finally, numerical results are put in Sec.~\ref{num} and we summarize in Sec.~\ref{summary}.

\section{Two-flavor Nambu--Jona-Lasinio model}\label{NJL}

\subsection{Formal derivations}\label{Formal}
To study charged pion superfluidity and rho superconductivity in the circumstances with finite isospin chemical potential $\mu_{\rm I}$ and magnetic field $B$, the original Lagrangian of the two-flavor Nambu--Jona-Lasinio (NJL) model~\cite{Klevansky:1992qe} can be extended to~\cite{Cao:2019res}
\bea
\mathcal{L} &=&\bar{\psi}\Bigl( i\gamma ^{\mu}D_{\mu\!}\!-\!m_0\!+\!\frac{\mu _I}{2}\gamma^0\tau _3 \Bigr)\psi \!+\!G_{\rm S}\!\left[(\bar{\psi}\psi )^2\!+\!(\bar{\psi}i\gamma _5\boldsymbol{\tau }\psi )^2\right] \nonumber\\
&&-G_{\rm V}\!\left[(\bar{\psi}\gamma ^{\mu}\tau ^a\psi )^2+(\bar{\psi}i\gamma ^{\mu}\gamma _5\tau ^a\psi )^2\right],
\eea
where $\psi=(u,d)^\mathrm{T}$ denotes the quark doublet, $m_0$ is the current quark mass, $\tau^a=(1,\boldsymbol{\tau})$ with $\boldsymbol{\tau}=(\tau_1,\tau_2,\tau_3)$ the Pauli matrices in flavor space, and $G_{\rm S}$ and  $G_{\rm V}$ are the coupling constants in the (pseudo-)scalar and (pseudo-)vector channels, respectively.
Here, the isospin chemical potential $\mu_{\rm I}$ is introduced as the energy mismatch between $u$ and $d$ quarks, and the magnetic effect is introduced through the covariant derivative $D_\mu=\partial_\mu+iqA_\mu$ where the electric-charge matrix is $q=\mathrm{diag}(q_u,q_d)$ in flavor space with $q_u=2e/3$ and $q_d=-e/3$. Without loss of generality, the vector potential will be chosen to be $A_\mu=(0,0,-Bx,0)$ corresponding to a constant magnetic field along $z$-direction, that is, $\mathbf{B}=B\hat{z}$.

To bilinearize the four-fermion interactions for later mean field calculations, auxiliary bosonic fields can be introduced~\cite{Cao:2019res}:
\begin{equation}
\begin{aligned}
\sigma &\equiv-2G_{\rm S}\,\bar{\psi}\psi, &
\boldsymbol{\pi} &\equiv-2G_{\rm S}\,\bar{\psi}i\gamma_5\boldsymbol{\tau}\psi,\\
V^{\mu a} &\equiv-2G_{\rm V}\,\bar{\psi}\gamma^{\mu}\tau^a\psi, &
A^{\mu a} &\equiv-2G_{\rm V}\,\bar{\psi}i\gamma^{\mu}\gamma_5\tau^a\psi .
\end{aligned}
\end{equation}
Then, by taking Hubbard-Stratonovich transformation with the help of these  auxiliary fields, the Lagrangian can be transformed to an equivalent form,
\bea
\mathcal{L}&=& \bar{\psi}\Bigl( i\slashed{D}\!-\!m_0\!-\!\sigma\!+\!\frac{\mu_{\rm I}}{2}\gamma^0\tau _3 \!-\!i\gamma_5\boldsymbol{\pi}\cdot\boldsymbol{\tau}\!+\!\slashed{V}^{a}\!\!\tau^a\!+\!\slashed{A}^{a}\!\tau^a i\gamma_5\Bigr)\psi \nonumber\\
&&-\frac{\sigma^2+\boldsymbol{\pi}^2}{4G_{\rm S}}+\frac{V^{\mu a}V_{\mu}^{ a}+A^{\mu a}A_{\mu}^{ a}}{4G_{\rm V}}
\eea
with $\slashed{X}\equiv X^\mu\gamma_\mu$. In the presence of magnetic field, it is more convenient to study with the physical meson fields which are eigenstates of electric charge. According to the following relations for the mesons we are interested in,
\bea
&&\pi^0 =\pi_3,\ \ \ \
\pi^{\pm} =\frac{1}{\sqrt{2}}(\pi_1\mp i\pi_2),\nonumber\\
&&\omega^\mu =V^{\mu 0}, \ \ 
\rho_{0}^{\mu} =V^{\mu 3}, \ \
\rho^{\pm\mu} =\frac{1}{\sqrt{2}}(V^{\mu 1}\mp iV^{\mu 2}),
\eea
the Lagrangian can be rewritten as
\bea
\mathcal{L} &=&\bar{\psi}\Bigl[ i\gamma^{\mu}\tilde{D}_{\mu}-m_0-\sigma+\frac{\mu_{\rm I}}{2}\gamma_0\tau_3
-i\gamma_5\bigl(\tau_3\pi^0\!+\!\tau_\pm\pi^\pm\bigr) \Bigr]\psi \nonumber\\
&&\!\!\!\!\!\!-\frac{\sigma^2\!+\!(\pi^0)^2\!+\!\pi^{\mp}\pi^{\pm}}{4G_{\rm S}} +\frac{(\omega^\mu)^2\!+\!(\rho_{0}^{\mu})^2
      \!+\!\rho_{\mu}^{\mp}\rho^{\pm\mu}
      \!+\!(A^{a\mu})^2}{4G_{\rm V}}\nonumber\\\label{HS}
\eea
with $\tau_\pm \equiv (\tau_1\pm i\tau_2)/\sqrt{2}$ and the modified covariant derivative
\begin{equation}
\tilde{D}_{\mu}= \partial_{\mu}+ i\Bigl(qA_{\mu}- \omega_{\mu}- \tau_3\rho_{0\mu}- \tau_\pm\rho^{\pm}_{\mu}- i\gamma_5\tau^a A_{\mu}^{a}\Bigr).
\end{equation}

In mean field approximation with only $\sigma$ condensation, the thermodynamic potential can be obtained by integrating over quark degrees of freedom as
\bea
\Omega={(m-m_0)^2\over4G_{\rm S}}+{i\over V_4}{\rm Tr}\ln {\cal G}^{-1}, 
\eea
where $m\equiv m_0+\langle\sigma\rangle$ is the dynamical mass , $G_{f}^{-1}\equiv i\gamma^{\mu}{D}_{\mu}-m+\frac{\mu_{\rm I}}{2}\gamma_0\tau_3$ is the mean-field inverse quark propagator, and the trace “${\rm Tr}$” are over the space-time coordinate, Dirac spinor, flavor, and color spaces. Then, the gap equation can be formally obtained by applying the minimum condition $\partial\Omega/\partial m=0$ as
\bea
{m-m_0\over 2G_{\rm S}}-{i\over V_4}{\rm Tr}\ {\cal G}=0,
\eea
where ${\cal G}={\rm diag}(G_{u},G_{d})$ is the quark propagator in flavor space.
It can be represented with the help of effective quark propagators $S_{u/d}(k)$ in energy momentum space as~\cite{Cao:2021rwx}
\bea
{m-m_0\over 2G_{\rm S}}+ \int\frac{d^4\mathbf{k}}{(2\pi)^4}{\rm tr}~{\cal S}(k)=0\label{GapE}
\eea
 with the trace ${\rm tr}$ over internal degrees of freedom and ${\cal S}\equiv{\rm diag}(S_{u},S_{d})$.

And the inverse meson propagators can be conveniently evaluated in random phase approximation (RPA) through~\cite{Cao:2015xja,Cao:2021rwx}:
\bea
\!\!\!\!\!\!D^{-1}_{SS}(y,x)&=&-{e^{-iq_{\rm S}\int_{x}^yA\cdot\di x}\over2G_{\rm S}}+{i\over V_4}{\rm Tr}~ {\cal G}\Gamma_{S^*} {\cal G}\Gamma_{S},\label{Sprp}\\
\!\!\!\!\!\!D^{-1}_{\bar{V}_{\mu}\bar{V}_{\nu}}(y,x)&=&{e^{-iq_{\rm V}\int_{x}^yA\cdot\di x}g_{{\mu}{\nu}}\over2G_{\rm V}}+{i\over V_4}{\rm Tr}~ {\cal G}\Gamma_{{\bar{V}_{\mu}}^*} {\cal G}\Gamma_{\bar{V}_{\nu}},\label{Vprp}
\eea
where the coupling vertices $\Gamma_{S/S^*}$ and $\Gamma_{\bar{V}_{\mu}/\bar{V}_{\mu}^*}$ can be read from \eqref{HS} as \cite{Cao:2019res}
\bea
&&\Gamma_{\sigma/\sigma^*}=-1,~\Gamma_{\pi^0/{\pi^0}^*}=-i\gamma^5\tau_3,~\Gamma_{\pi_\pm}=-i\gamma^5\tau_\pm,\nonumber\\
&&\Gamma_{\bar{\omega}_\mu/\bar{\omega}_\mu^*}=\bar{\gamma}_\mu^\pm,~\Gamma_{\bar{\rho}_{0\mu}/\bar{\rho}_{0\mu}^*}=\bar{\gamma}_\mu^\pm\tau_3,~\Gamma_{\bar{\rho}_{\pm\mu}}=\bar{\gamma}_\mu^\pm\tau_\pm.
\eea
Note that $\bar{V}_{\mu}/\bar{V}_{\mu}^*\equiv({V}_{0},{{V}_{1}\mp i{V}_{2}\over \sqrt 2},{{V}_{1}\pm i{V}_{2}\over \sqrt 2},V_3)$ are the spin eigenstates with the spatial components $\bar{V}_{1}, \bar{V}_{2}$ and $\bar{V}_{3}$ corresponding to spin components $S_z=1,-1$ and $0$ along ${\bf B}$, and the corresponding gamma matrices are $\bar{\gamma}^{\pm}_\mu\equiv(\gamma_0,{\gamma_1\pm i\gamma_2\over \sqrt{2}},{\gamma_1\mp i\gamma_2\over \sqrt{2}},\gamma_3)$. 

It is very important to point out that: For nonlocal meson propagators, the Schwinger phases should be compensated for the charged mesons in order to keep the theory gauge invariance in external EM field~\cite{Cao:2015xja}, see the Wilson lines in the first terms of Eq.\eqref{Sprp} and Eq.\eqref{Vprp} with the integral along a straight line. Then, their masses could be evaluated in energy momentum space by dropping the gauge dependent Schwinger phases, that is, by requiring $D^{-1}(p_0,{\bf p}={\bf 0})=0$ for
\bea
\!\!\!\!\!\!\!\!D^{-1}_{SS}(p)&\equiv&{1\over2G_{\rm S}}+\Pi_{SS}(p)\nonumber\\
&=&\int\di^4x~e^{-ip\cdot (y-x)}e^{iq_{\rm S}\int_{x}^yA\cdot\di x}D^{-1}_{SS}(y,x),\label{SprpP}\\
\!\!\!\!\!\!\!\!D^{-1}_{\bar{V}_{\mu}\bar{V}_{\mu}}(p)&\equiv&{1\over2G_{\rm V}}+\Pi_{\bar{V}_{\mu}\bar{V}_{\mu}}(p)\nonumber\\
&=&\int\di^4x~e^{-ip\cdot (y-x)}e^{iq_{\rm V}\int_{x}^yA\cdot\di x}D^{-1}_{\bar{V}_{\mu}\bar{V}_{\mu}}(y,x).\label{VprpP}
\eea
Actually, it is equivalent to derive these inverse propagators directly with the help of the effective quark propagators $S_{f}({k})$ defined in energy momentum space, that is,
\bea
\Pi_{SS}(p)&=&\int\frac{d^4\mathbf{k}}{(2\pi)^4}{\rm tr}~ {\cal S}(k)\Gamma_{S^*} {\cal S}(p+k)\Gamma_{S},\nonumber\\
\Pi_{\bar{V}_{\mu}\bar{V}_{\mu}}(p)&=&\int\frac{d^4\mathbf{k}}{(2\pi)^4}{\rm tr}~{\cal S}(k)\Gamma_{{\bar{V}_{\mu}}^*} {\cal S}(p+k)\Gamma_{\bar{V}_{\nu}}.\label{SEn}
\eea
To explore new homogeneous phases, the quadratic Ginzburg-Landau expansion coefficients are related to the self energies \eqref{SEn} as the following
\bea
 \mathcal{A}_{\rm SS}={1\over4G_{\rm S}}\!+\!{\Pi_{\rm SS}(0)\over 2}, \mathcal{A}_{\bar{V}_{\mu}\bar{V}_{\mu}}={1\over4G_{\rm V}}\!+\!{\Pi_{\bar{V}_{\mu}\bar{V}_{\mu}}(0)\over 2}.\label{GLEC}
\eea

\subsection{Explicit evaluations and regularizations}\label{Explicit}
In this section, we are going to derive the explicit forms for the gap equation \eqref{GapE} and the quadratic Ginzburg-Landau coefficients \eqref{GLEC} with the help of the explicit forms of the quark propagators. For a given uniform magnetic field along $z$-direction, it is well known that the fermion propagator can be written as a gauge-covariant phase factor multiplying a translationally invariant kernel in the coordinate space. So, the quark propagators can be given explicitly as \cite{Schwinger:1951nm}

\begin{widetext}
\begin{equation}
\begin{aligned}
G_f(x,y) &= \exp\!\left[-iq_f\int_y^x \bar{A}_f^\mu\,dx_\mu\right]\,S_f(x-y)\,,
\\
S_f(x) &= -i\int_0^{\infty}\frac{ds}{16(\pi s)^2}\,
\exp\!\left\{-i\left[m^2s+\frac{1}{4s}\Big(x_{0}^{2}-x_{3}^{2}-\mathbf{x}_{\bot}^{2}(q_fBs)\cot(q_fBs)\Big)\right]\right\}
\\
&\quad \times \left[m+\frac{1}{2s}\Big(\gamma_0x_0-\gamma_3x_3-(q_fBs)\big((\gamma_1x_1+\gamma_2x_2)\cot(q_fBs)+\gamma_2x_1-\gamma_1x_2\big)\Big)\right] (q_fBs)\Big[\cot(q_fBs)-\gamma_1\gamma_2\Big],
\end{aligned}
\end{equation}
where $\mathbf{x}_{\bot}^{2}=x_{1}^{2}+x_{2}^{2}$ and the isospin chemical potential is introduced through the effective potential $\bar{A}_{u/d}^{\mu}=\Big(\mp \frac{\mu _I}{2q_f},0,0,0\Big)+A^{\mu}$. Note that the integral in the effective Schwinger phase is along a straight line from $y$ to $x$. Getting rid of the Schwinger phase with respect to $A^{\mu}$, the left terms are translationally invariant thus can be transformed to energy momentum space. Usually, it is legal to rotate the proper time from $s$ to $-i\,s$ for a constant magnetic field, so the quark propagators in Euclidean energy momentum space can be given as \cite{Cao:2015xja}
\bea
S_{f}^{E}(k)
&=&-i\int_{0}^{\infty}ds\,
\exp\!\left[-s\left(m^2+k_{4}^{f\,2}+k_{3}^{2}+\mathbf{k}_{\bot}^{2}\frac{\tanh(q_fBs)}{q_fBs}\right)\right]
\nonumber\\
&&\quad \times
\left(-\gamma_{\mu}k_{\mu}+m+i(\gamma_1k_2-\gamma_2k_1)\tanh(q_fBs)\right)
\left(1+i\gamma_1\gamma_2\tanh(q_fBs)\right)\label{SES}
\eea
with $\mathbf{k}_{\bot}^{2}=k_{1}^{2}+k_{2}^{2}$. At finite temperature and isospin density, the imaginary time components are defined as $k_{4}^{u/d}\equiv\omega _n\pm i\frac{\mu _I}{2}$ with the Matsubara frequency $\omega_n=(2n+1)\pi T\ (n\in \mathcal{Z})$.
As we can see, the large-$s$ convergence is controlled by $\mathrm{Re}\!\left[m^2+(k_4^f)^2+k_3^2+\cdots\right]$ which is positive definite without $\mu _I$. However, when $\mu _I$ is present and sufficiently large, $(k_4^f)^2=\omega_n^2-\mu_{\rm I}^2/4+i\omega_n\mu_{\rm I}$ can render the real part negative for some modes, so the proper-time rotation is illegal and the propagator is invalid.

The invalidity comes from the introduction of proper time in Schwinger's formalism but can be well avoided if one rewrites the propagator in the Landau-level representation. By utilizing the identity $\tanh(x)=1-\frac{2e^{-2x}}{1+e^{-2x}}$ and the generating function of Laguerre polynomials,
\bea
(1-z)^{-(\alpha+1)}\exp\!\left(\frac{xz}{z-1}\right)=\sum_{n=0}^{\infty}L_n^{\alpha}(x)\,z^n,
\eea
the quark propagators \eqref{SES} can be rewritten as \cite{Chodos:1990vv, Miransky:2015ava}
\bea
	S_{f}^{E}(k)&=&-i\exp \left( -\frac{\mathbf{k}_{\bot}^{2}}{\left| q_fB \right|} \right) \sum_{n=0}^{\infty}{\left( -1 \right) ^n\frac{D_n(q_fB,k)}{\left( k_{4}^{f} \right) ^2+k_{3}^{2}+m^2+2\left| q_fB \right|n}}\nonumber\\
	D_n(q_fB,k)&=&\left( m-k_{4}^{f}\gamma _4-k_3\gamma _3 \right) \left\{ \left[ 1+i\gamma _1\gamma _2\mathrm{sign(}q_fB) \right] L_n\left( 2\frac{\mathbf{k}_{\bot}^{2}}{\left| q_fB \right|} \right) -\left[ 1-i\gamma _1\gamma _2\mathrm{sign(}q_fB) \right] L_{n-1}\left( 2\frac{\mathbf{k}_{\bot}^{2}}{\left| q_fB \right|} \right) \right\}\nonumber\\
	&&+4\left( k_1\gamma _1+k_2\gamma _2 \right) L_{n-1}^{1}\left( 2\frac{\mathbf{k}_{\bot}^{2}}{\left| q_fB \right|} \right)\label{SEL}
\eea
where $L_n(x)\equiv L_n^{0}(x)$ and $L_{-1}^{\alpha}(x)=0$. As we can see from the denominator, $n$ is nothing else but the Landau levels; and just like the case without magnetic field, such a from is valid to study the effect of however large $\mu _I$. In the following, we will use proper-time representation for the cases when only magnetic field is involved as the terms can be renormalized conveniently, but refer to the Landau representation for the cases when both magnetic field and isospin chemical potential are involved to keep the results valid.

Plugging the propagators \eqref{SEL} into the gap equation \eqref{GapE} and performing the Matsubara frequency summation, we have
\bea
0&=&\frac{m-m_0}{2G_{\rm S}}
+\frac{N_c}{\pi}\sum_{f=u,d}|q_fB|\sum_{n=0}^{\infty}\left(1-\frac{\delta_{n0}}{2}\right)
\int\frac{dk_3}{2\pi}\frac{m}{E_{f}}
\left[\frac{1}{e^{(E_{f}+\mu_{\rm I}/2)/T}+1}+\frac{1}{e^{(E_{f}-\mu_{\rm I}/2)/T}+1}-1\right],
\eea
where the Landau-level dispersion is $E_{f}\equiv\sqrt{k_3^2+m^2+2|q_fB|\,n}$. The solely $B$-dependent term in the integral is divergent and can be regularized with the help of proper-time representation, then the gap equation becomes \cite{Cao:2021rwx}
\bea
0&=&\frac{m-m_0}{2G}
-4N_c\int^{\Lambda}\frac{d^3\mathbf{k}}{(2\pi)^3}\frac{m}{\sqrt{\mathbf{k}^2+m^2}}
-\frac{N_cm}{4\pi^2}\sum_{f=u,d}\int_0^{\infty}\frac{ds}{s^2}e^{-m^2s}\left(\frac{q_fBs}{\tanh(q_fBs)}-1\right)
\nonumber\\
&&\quad+\frac{N_c}{\pi}\sum_{f=u,d}|q_fB|\sum_{n=0}^{\infty}\left(1-\frac{\delta_{n0}}{2}\right)
\int\frac{dk_3}{2\pi}\frac{m}{E_{f}}
\left[\frac{1}{e^{(E_{f}+\mu_{\rm I}/2)/T}+1}+\frac{1}{e^{(E_{f}-\mu_{\rm I}/2)/T}+1}\right].\label{gap}
\eea

Similarly, the quadratic Ginzburg-Landau coefficient of $\pi^+$ and the lowest energy spin state $\bar{\rho}_1^{+}$ can be evaluated according to \eqref{GLEC} as
\bea
 \mathcal{A}_{\pi^+\pi^+} &=&{1\over4G_{\rm S}}-N_c\,T\sum_{n}\int\frac{d^3\mathbf{k}}{(2\pi)^3}
\mathrm{tr} \left[ S_u^{E}(k_4^{u},\mathbf{k})\,i\gamma_5\, S_d^{E}(k_4^{d},\mathbf{k})\,i\gamma_5 \right]\nonumber\\
&=&{1\over4G_{\rm S}}+8N_c\,T\sum_{n}\int\frac{d^3\mathbf{k}}{(2\pi)^3}
\sum_{n_1,n_2=0}^{\infty}
\frac{(-1)^{n_1+n_2}e^{-\frac{9\mathbf{k}_{\perp}^{2}}{2eB}}}{\big[(k_4^{u})^2+E_{u}^2\big]\big[(k_4^{d})^2+E_{d}^2\big]} \Bigg\{
\big(k_3^2+k_4^{u}k_4^{d}+m^2\big)
\left[
L_{n_1-1}\!\left(\frac{3\mathbf{k}_{\perp}^{2}}{eB}\right)
L_{n_2}\!\left(\frac{6\mathbf{k}_{\perp}^{2}}{eB}\right)\right.\nonumber\\
&& \left. +L_{n_1}\!\left(\frac{3\mathbf{k}_{\perp}^{2}}{eB}\right)
L_{n_2-1}\!\left(\frac{6\mathbf{k}_{\perp}^{2}}{eB}\right)
\right]
-8\mathbf{k}_{\perp}^{2}\,
L_{n_1-1}^{1}\!\left(\frac{3\mathbf{k}_{\perp}^{2}}{eB}\right)
L_{n_2-1}^{1}\!\left(\frac{6\mathbf{k}_{\perp}^{2}}{eB}\right)
\Bigg\},\\
 \mathcal{A}_{\bar{\rho}_1^{+}\bar{\rho}_1^{+}} &=&{1\over4G_{\rm V}} -N_c\,T\sum_{n}\int\frac{d^3\mathbf{k}}{(2\pi)^3}
\mathrm{tr} \left[ S_u^{E}(k_4^{u},\mathbf{k})\,\bar{\gamma}_1^{+}\, S_d^{E}(k_4^{d},\mathbf{k})\,\bar{\gamma}_1^{-} \right]\nonumber\\
&=&{1\over4G_{\rm V}}-16N_c\,T\sum_{n}\int\frac{d^3\mathbf{k}}{(2\pi)^3}
e^{-\frac{9\mathbf{k}_{\perp}^{2}}{2eB}}
\sum_{n_1,n_2=0}^{\infty}(-1)^{n_1+n_2}
\frac{k_3^2+k_4^{u}k_4^{d}+m^2}{\big[(k_4^{u})^2+E_u^2\big]\big[(k_4^{d})^2+E_d^2\big]}L_{n_1}\!\left(\frac{3\mathbf{k}_{\perp}^{2}}{eB}\right)
L_{n_2}\!\left(\frac{6\mathbf{k}_{\perp}^{2}}{eB}\right)
\eea
with $eB$ positive. In the following, we will regularize $ \mathcal{A}_{\pi^+\pi^+}$ and $ \mathcal{A}_{\bar{\rho}_1^{+}\bar{\rho}_1^{+}} $ one by one.

\subsubsection{The Ginzburg-Landau coefficient of $\pi^+$}\label{Pi}
To streamline the subsequent evaluation of the Matsubara sum and the remaining longitudinal momentum integral, it is convenient to treat first the integration with respect to the transverse momentum, $\mathbf{k}_{\perp}=(k_1,k_2)$. For $\pi^+$, we isolate the $\mathbf{k}_{\perp}$-dependent pieces and define
\bea
I_{\pi^+,\perp}&\equiv&
\int\frac{d^2\mathbf{k}_{\perp}}{(2\pi)^2}
\sum_{n_1,n_2=0}^{\infty}(-1)^{n_1+n_2}
\exp\!\left(-\frac{9\mathbf{k}_{\perp}^{2}}{2eB}\right)
\Bigg\{
\left(k_3^{2}+k_4^{u}k_4^{d}+m^2\right)
\Big[
L_{n_1-1}\!\left(\frac{3\mathbf{k}_{\perp}^{2}}{eB}\right)
L_{n_2}\!\left(\frac{6\mathbf{k}_{\perp}^{2}}{eB}\right) \nonumber\\
&&\ \ \ \ \ \ \ \ \ \ \ \ \ \ \ \ \ \ \ +
L_{n_1}\!\left(\frac{3\mathbf{k}_{\perp}^{2}}{eB}\right)
L_{n_2-1}\!\left(\frac{6\mathbf{k}_{\perp}^{2}}{eB}\right)
\Big]
-8\mathbf{k}_{\perp}^{2}\,
L_{n_1-1}^{1}\!\left(\frac{3\mathbf{k}_{\perp}^{2}}{eB}\right)
L_{n_2-1}^{1}\!\left(\frac{6\mathbf{k}_{\perp}^{2}}{eB}\right)
\Bigg\}.
\eea
At this stage, all dependence on the external magnetic field in the transverse plane is encoded in the Gaussian factor and the Laguerre polynomials, while the remaining variables, $k_4^{u,d},k_3$, and $m$, play roles of overall prefactors. We now convert the transverse integral to polar coordinates $(k_\perp,\theta)$ and introduce the dimensionless variable $x\equiv \frac{3\mathbf{k}_{\perp}^{2}}{eB}$,  then $\exp\!\left(-9\mathbf{k}_{\perp}^{2}/(2eB)\right)=e^{-3x/2}$ and the Laguerre arguments become $x$ and $2x$ with respect to $d$ and $u$ quarks. Consequently, the transverse part takes the form
\bea
I_{\pi^+,\perp}
\!&=&\!\frac{eB}{12\pi}\!\int_{0}^{\infty}\!\!\!\!\!dx\,e^{-{3\over2}x}\!
\!\!\!\!\!\sum_{n_1,n_2=0}^{\infty}\!\!\!(-1)^{n_1\!+n_2}\!
\left\{
\left(k_3^{2}\!+\!k_4^{u}k_4^{d}\!+\!m^2\right)
\!\!\Big[
L_{n_1\!-\!1}\!(x)L_{n_2}\!(2x)\!+\!L_{n_1}\!(x)L_{n_2-\!1}\!(2x)
\Big] 
\!\!-\!8\frac{eB}{3}x\,L_{n_1\!-\!1}^{1}\!(x)L_{n_2-\!1}^{1}\!(2x)
\right\}.\nonumber\\
\eea

With the help of the standard integral identities for generalized Laguerre polynomials~\cite{Gradshteyn1980}, 
\bea
\int_{0}^{\infty}\!\!\!dx\,e^{-bx}\,x^{\alpha}\,L_{n}^{\alpha}(\lambda x)\,L_{m}^{\alpha}(\mu x)
=\frac{\Gamma(m\!+\!n\!+\!\alpha\!+\!1)}{m!\,n!}
\frac{(b\!-\!\lambda)^n(b\!-\!\mu)^m}{b^{m+n+\alpha+1}}\,
{}_2F_1\!\left(
-m,-n;\,-m-n-\alpha;\,
\frac{b(b\!-\!\lambda\!-\!\mu)}{(b\!-\!\lambda)(b\!-\!\mu)}
\right),
\eea
the integral over $x$ can be completed to find
\bea
I_{\pi^+,\perp}
\!&=&\!\frac{eB}{12\pi}\sum_{n_1,n_2=0}^{\infty}\!\!\!(-1)^{n_1\!+n_2}
\left\{
\left(k_3^{2}\!+\!k_4^{u}k_4^{d}\!+\!m^2\right)
\left[
C'(n_1-1,n_2)\!+\!C'(n_1,n_2-1)
\right] 
-8\frac{eB}{3}C(n_1,n_2)
\right\}
\eea
with the coefficients defined as
\bea
    C(n_1,n_2)&\equiv&
\int_{0}^{\infty}\!\!\!dx\,x\,e^{-{3\over2}x}\,L_{n_1\!-\!1}^{1}(x)\,L_{n_2-\!1}^{1}(2x)
=-\frac{4(-1)^{n_2}}{3^{n_1\!+n_2}}\frac{\Gamma(n_1\!+\!n_2)}{(n_1\!-\!1)!\,(n_2\!-\!1)!}\,
{}_2F_1\!\left(1\!-\!n_2,1\!-\!n_1;\,1\!-\!n_1\!-\!n_2;\,9\right),\\
C'(n_1,n_2)
&\equiv&
\int_{0}^{\infty}\!\!\!dx\,e^{-{3\over2}x}\,L_{n_1}(x)\,L_{n_2}(2x) 
=
\frac{2(-1)^{n_2}}{3^{n_1\!+n_2\!+\!1}}\,
\frac{\Gamma(n_1\!+\!n_2+\!1)}{n_1!\,n_2!}\,
{}_2F_1\!\left(-n_2,-n_1;\,-n_1\!-\!n_2;\,9\right).
\eea
By applying the identity $\left( c-a-b \right) {}_2F_1(a,b;c;z)+a\left( 1-z \right) {}_2F_1(a+1,b;c;z)-\left( c-b \right) {}_2F_1(a,b-1;c;)z=0$~\cite{Abramowitz1972}, we find $C'(n_1-1,n_2)+C'(n_1,n_2-1)=4\left({1\over n_2}-{1\over n_1}\right) C(n_1,n_2)$. Thus,
\bea
I_{\pi^+,\perp}
\!&=&\!\frac{eB}{3\pi}\sum_{n_1,n_2=0}^{\infty}\!\!\!(-1)^{n_1\!+n_2}
\left\{
\left(k_3^{2}\!+\!k_4^{u}k_4^{d}\!+\!m^2\right)
\left({1\over n_2}-{1\over n_1}\right)
-2\frac{eB}{3}
\right\}C(n_1,n_2),
\eea
and it follows that 
\bea
 \mathcal{A}_{\pi^+\pi^+} ={1\over4G_{\rm S}}\!+\!\frac{2N_c}{3\pi}eB\!\sum_{n_1,n_2}^{\infty}\!(-1)^{n_1+n_2}
\int\frac{dk_3}{2\pi}\,T\sum_{n}\,
\left\{
\frac{\tfrac{4}{n_2}}{(k_4^{u})^2\!+\!E_u^2}
\!-\!\frac{\tfrac{4}{n_1}}{(k_4^{d})^2\!+\!E_d^2}
\!+\!\frac{i\mu_{\rm I}\left(\tfrac{4}{n_2}k_4^{d}\!+\!\tfrac{4}{n_1}k_4^{u}\right)}
{\big[(k_4^{u})^2\!+\!E_u^2\big]\big[(k_4^{d})^2\!+\!E_d^2\big]}
\right\}\,C(n_1,n_2).\nonumber\\
\eea

Then, we can complete the summation over fermion Matsubara frequency and get
\bea
 \mathcal{A}_{\pi^+\pi^+} ={1\over4G_{\rm S}}+\Pi_{\pi^+\pi^+}^{B}+\Pi_{\pi^+\pi^+}^{B,\mu _I}+\Pi_{\pi^+\pi^+}^{B,\mu _I,T},
\eea
where the solely $B$-dependent part can be given in proper-time representation as
\bea
\Pi_{\pi^+\pi^+}^{B}
&=&-{N_c\over 4\pi^2}\int_{0}^{\infty}\!\!\!ds\int_{0}^{\infty}{e^{-m^2(s+t)}dt\over s+t}\,
\left( \frac{\tanh\!\left(q_uBs\right)}{q_uB}\!+\!\frac{\tanh\!\left(q_dBt\right)}{q_dB}\right)^{-1} \Bigg[
\left(m^2\!+\!\frac{1}{s\!+\!t}\right)\left(1\!-\!\tanh\!\left(q_uBs\right)\tanh\!\left(q_dBt\right)\right)\nonumber\\
&&+\left(\frac{\tanh\!\left(q_uBs\right)}{q_uB}+\frac{\tanh\!\left(q_dBt\right)}{q_dB}\right)^{-1} \left(1-\tanh^2\!\left(q_uBs\right)\right)\left(1-\tanh^2\!\left(q_dBt\right)\right)
\Bigg],
\eea
 the  $B$ and $\mu _I$ dependent part is
\bea
\Pi _{\pi^+\pi^+}^{B,\mu _I}
=-\frac{2N_c}{3\pi}\sum_{n_1,n_2=0}^{\infty}g_{\pi}(eB,n_1,n_2),\ \ g_{\pi}(eB,n_1,n_2)\equiv eB\int\frac{dk_3}{2\pi}\,
\frac{2(-1)^{n_1+n_2}\mu _{I}^{2}\left( n_2E_u-n_1E_d \right)}{E_uE_dn_1n_2\left[ \left( E_u+E_d \right) ^2-\mu _{I}^{2} \right]}\,
C(n_1,n_2),
\eea
and the thermal part is
\bea
\Pi _{\pi^+\pi^+}^{B,\mu _I,T}
&=&\frac{4N_c}{3\pi}eB\sum_{n_1,n_2=0}^{\infty}(-1)^{n_1+n_2}\sum_{t=\pm}\int\frac{dk_3}{2\pi}\,
 \left\{
\left[ -\frac{1}{n_2E_u}+\frac{t\,\mu _I\left( E_u(n_1+n_2)+t\,n_1\mu _I \right)}{n_1n_2E_u\left[ (E_u+t\,\mu _I)^2-E_{d}^{2} \right]} \right]
\frac{1}{e^{\frac{2E_u+t\,\mu _I}{2T}}+1}\right.\nonumber\\
&&\left.\quad\quad\quad\quad\quad\quad\quad\quad\quad+
\left[ \frac{1}{n_1E_d}-\frac{t\,\mu _I\left( E_d(n_1+n_2)+t\,n_2\mu _I \right)}{n_1n_2E_d\left[ (E_d+t\,\mu _I)^2-E_{u}^{2} \right]} \right]
\frac{1}{e^{\frac{2E_d+t\,\mu _I}{2T}}+1}
\right\}\,C(n_1,n_2).
\eea

The non-thermal parts are divergent thus need regularization. If we take $B\rightarrow 0$ for those parts, we must get the divergent part of the result with only finite $\mu_{\rm I}$, which was well known previously. On the other hand, the magnetic field itself usually would not induce divergence and the $B$-dependent parts can be separated out by performing subtractions~\cite{Cao:2021rwx}. Regarding $\Pi_{\pi^+\pi^+}^{B}$ and $\Pi_{\pi^+\pi^+}^{B,\mu _I}$, we have
\bea
\Delta\Pi_{\pi^+\pi^+}^{B}
\!&=&\!-{N_c\over 4\pi^2}\int_{0}^{\infty}\!\!\!ds\int_{0}^{\infty}\!\!{e^{-m^2(s+t)}dt\over s+t}\,
\left\{\left( \frac{\tanh\!\left(q_uBs\right)}{q_uB}\!+\!\frac{\tanh\!\left(q_dBt\right)}{q_dB}\right)^{-1} \Bigg[
\left(m^2\!+\!\frac{1}{s\!+\!t}\right)\left(1\!-\!\tanh\!\left(q_uBs\right)\tanh\!\left(q_dBt\right)\right)\right.\nonumber\\
&&\left.+\left(\frac{\tanh\!\left(q_uBs\right)}{q_uB}+\frac{\tanh\!\left(q_dBt\right)}{q_dB}\right)^{-1} \left(1-\tanh^2\!\left(q_uBs\right)\right)\left(1-\tanh^2\!\left(q_dBt\right)\right)
\Bigg]-{2+m^2(s+t)\over(s+t)^2}\right\},\label{DPiB}\\
\Delta \Pi_{\pi^+\pi^+}^{B,\mu_{\rm I}}
&=&-\frac{2N_c}{3\pi}\left[
\sum_{n_1,n_2=0}^{N_1,N_2} g_\pi(eB,n_1,n_2)
-\sum_{n_1',n_2'=0}^{N_1',N_2'} g_\pi(eB',n_1',n_2')
\right],\label{DPiBM}
\eea
where $N_1, N_2, N_1'$ and $N_2'$ are the upper limits of $n_1, n_2, n_1'$ and $n_2'$, and $B'$ is set to be small to mimic the vanishing magnetic limit in the second term. Note that a $B$-independent ultraviolet cutoff $\Lambda^2$ should be introduced to regularize the Landau eigenenergies, in $\Delta \Pi_{\pi\pi}^{B,\mu_{\rm I}}$, that is, $|q_uBn_1|, |q_dBn_2|,|q_uB'n_1'|, |q_dB'n_2'|\leq\Lambda^2$. In other words, if we take $b\equiv B/B'\in\mathcal{N}$, then the upper limits are related to each other as $N_2=2N_1, N_2'=2N_1'$, and $N_1'=bN_1$. For comparison, a naive choice of the same cutoff to the Landau levels, $N_1=N_2=N_1'=N_2'$, can never render $\Delta \Pi_{\pi\pi}^{B,\mu_{\rm I}}$ convergent, since 
it is the compact Landau eigenenergies that are truly involved in $g_\pi(eB,n_1,n_2)$. For two arbitrary magnetic fields, $B_1$ and $B_2$, we can choose a small enough $B'$ such that $b_i\equiv B_i/B'\in\mathcal{N}\ (i=a,b)$; then the $B'$-dependent terms cancel out in $\Delta \Pi_{\pi^+\pi^+}^{B_1,\mu_{\rm I}}
|_{N_1\rightarrow b_2 N}-\Delta \Pi_{\pi^+\pi^+}^{B_2,\mu_{\rm I}}|_{N_1\rightarrow b_1 N}$ and $\sum_{n_1,n_2=0}^{b_2N,2b_2N} g_\pi(eB_1,n_1,n_2)
-\sum_{n_1,n_2=0}^{b_1N,2b_1N} g_\pi(eB_1,n_1,n_2)$ is found to be convergent. Note that our regularization scheme is consistent with the proper-time regularization scheme, that is, when we subtract the magnetic dependent integrands by their vanishing magnetic field limits in $2+1$ dimensions, the lower limit of proper-time integral can usually be taken to zero~\cite{Ebert:1999ht}. 

Eventually, by collecting the $\mu_{\rm I}$-dependent divergent parts, which have been subtracted in \eqref{DPiB} and \eqref{DPiBM}, and regularizing them with three-momentum cutoff $\Lambda$, we obtain a finite Ginzburg-Landau coefficient,
\bea
 \mathcal{A}_{\pi^+\pi^+} ={1\over4G_{\rm S}}-2N_c\int^{\Lambda}\frac{d^3\mathbf{k}}{(2\pi)^3}\,
\frac{E_{\mathbf{k}}}{E_{\mathbf{k}}^{2}-(\mu_{\rm I}/2)^2}+\Delta\Pi_{\pi^+\pi^+}^{B}+\Delta\Pi_{\pi^+\pi^+}^{B,\mu _I}+\Pi_{\pi^+\pi^+}^{B,\mu _I,T}.\label{Api}
\eea

\subsubsection{The Ginzburg-Landau coefficient of $\bar{\rho}_1^{+}$}\label{Rho}
The regularization of the Ginzburg-Landau coefficient of $\bar{\rho}_1^{+}$ follows the same scheme but is simpler. The $\mathbf{k}_{\perp}$-dependent pieces can be isolated as
\bea
I_{\bar{\rho}_1^{+},\perp}(n_1,n_2)\equiv
\int\frac{d^2\mathbf{k}_{\perp}}{(2\pi)^2}
\exp\!\left(-\frac{9\mathbf{k}_{\perp}^{2}}{2eB}\right)
L_{n_1}\!\left(\frac{3\mathbf{k}_{\perp}^{2}}{eB}\right)
L_{n_2}\!\left(\frac{6\mathbf{k}_{\perp}^{2}}{eB}\right).
\eea
Then, by converting the transverse integral to polar coordinates $(k_\perp,\theta)$, it can be evaluated to be
\begin{equation}
I_{\bar{\rho}_1^{+},\perp}(n_1,n_2)=\frac{eB}{12\pi}C'(n_1,n_2).
\end{equation}
And the Ginzburg-Landau coefficient follows as
\bea
\mathcal{A}_{\bar{\rho}_1^{+}\bar{\rho}_1^{+}}
&=&{1\over4G_{\rm V}}-{4N_c\over 3\pi}eB \sum_{n_1,n_2=0}^{\infty}(-1)^{n_1+n_2}T\sum_{n}\int\frac{d{k}_3}{2\pi}
\frac{k_3^2+k_4^{u}k_4^{d}+m^2}{\big[(k_4^{u})^2+E_u^2\big]\big[(k_4^{d})^2+E_d^2\big]}C'(n_1,n_2)
\eea
By completing the summation over fermion Matsubara frequency, we have
\bea
 \mathcal{A}_{\bar{\rho}_1^{+}\bar{\rho}_1^{+}}={1\over4G_{\rm V}}+\Pi_{\bar{\rho}_1^{+}\bar{\rho}_1^{+}}^{B}+\Pi_{\bar{\rho}_1^{+}\bar{\rho}_1^{+}}^{B,\mu _I}+\Pi_{\bar{\rho}_1^{+}\bar{\rho}_1^{+}}^{B,\mu _I,T},
\eea
where the solely $B$-dependent part can be given in proper-time representation as
\bea
\Pi_{\bar{\rho}_1^{+}\bar{\rho}_1^{+}}^{B}
&=&-\frac{N_c}{8\pi ^2}\int_0^{\infty}ds\int_{-1}^1 du\,
e^{-m^2s}\left[
\frac{\tanh \!\left( \frac{eBs}{3}(1+u)\right)}{\frac{2}{3}eB}
+\frac{\tanh \!\left( \frac{eBs}{6}(1-u)\right)}{\frac{1}{3}eB}
\right] ^{-1}
\left( m^2+\frac{1}{s} \right) \nonumber\\
&&\ \ \ \ \ \ \ \ \ \ \ \ \left[ 1+\tanh \!\left( \frac{eBs}{6}(1-u) \right) \right]
\left[ 1+\tanh \!\left( \frac{eBs}{3}(1+u) \right) \right],
\eea
 the  $B$ and $\mu _I$ dependent part is
\bea
\!\!\!\!\Pi _{\bar{\rho}_1^{+}\bar{\rho}_1^{+}}^{B,\mu _I}
=-\frac{2N_c}{3\pi}\!\!\!\sum_{n_1,n_2=0}^{\infty}\!\!g_{\rho}(eB,n_1,n_2),\ \ g_{\rho}(eB,n_1,n_2)\equiv eB\int_{-\infty}^{+\infty}\frac{dk_3}{2\pi}\,
\frac{(-1)^{n_1+n_2}\mu_{\rm I}^2\big(k_3^2\!+\!m^2\!+\!E_uE_d\big)C'(n_1,n_2)}
{E_uE_d\,(E_u+E_d)\big[(E_u+E_d)^2-\mu_{\rm I}^2\big]}\,,
\eea
and the thermal part is
\bea
\Pi _{\bar{\rho}_1^{+}\bar{\rho}_1^{+}}^{B,\mu _I,T}
&=&-\frac{2N_c}{3\pi}eB\sum_{n_1,n_2=0}^{\infty}(-1)^{n_1+n_2}\sum_{t=\pm}\int\frac{dk_3}{2\pi}
\left[
\frac{1}{E_u}
\frac{k_{3}^{2}+m^2-E_u\left( E_u+t\,\mu _I \right)}{\left( E_u+t\,\mu _I \right) ^2-E_{d}^{2}}\,
\frac{1}{e^{\frac{2E_u+t\,\mu _I}{2T}}+1}
\right.\nonumber
\\
&&\qquad\qquad\qquad\left.+
\frac{1}{E_d}
\frac{k_{3}^{2}+m^2-E_d\left( E_d-t\,\mu _I \right)}{\left( E_d-t\,\mu _I \right) ^2-E_{u}^{2}}\,
\frac{1}{e^{\frac{2E_d-t\,\mu _I}{2T}}+1}
\right]
\,C'(n_1,n_2).
\eea

Again, the non-thermal parts are divergent thus need regularization. Different from $\pi^+$, a finite spin is involved  for $\bar{\rho}_1^{+}$, so the spin-magnetic field coupling should also be regularized~\cite{Cao:2019res} and we have
\bea
\Delta\Pi_{\bar{\rho}_1^{+}\bar{\rho}_1^{+}}^{B}
\!&=&\!-\frac{N_c}{8\pi ^2}\int_0^{\infty}ds\int_{-1}^1 du\,
e^{-m^2s}\left\{\left[
\frac{\tanh \!\left( \frac{eBs}{3}(1+u)\right)}{\frac{2}{3}eB}
+\frac{\tanh \!\left( \frac{eBs}{6}(1-u)\right)}{\frac{1}{3}eB}
\right] ^{-1}
\left( m^2+\frac{1}{s} \right)\right. \nonumber\\
&&\ \ \ \ \ \ \ \ \ \ \ \ \left.\left[ 1+\tanh \!\left( \frac{eBs}{6}(1-u) \right) \right]
\left[ 1+\tanh \!\left( \frac{eBs}{3}(1+u) \right) \right]-{1\over s}\left( m^2+\frac{1}{s} \right)\left( 1+\frac{eBs}{2} \right)\right\},\label{DRhoB}\\
\Delta \Pi_{\bar{\rho}_1^{+}\bar{\rho}_1^{+}}^{B,\mu_{\rm I}}
&=&-\frac{2N_c}{3\pi}\left[
\sum_{n_1,n_2=0}^{N_1,N_2} g_\rho(eB,n_1,n_2)
-\sum_{n_1',n_2'=0}^{N_1',N_2'} g_\rho(eB',n_1',n_2')
\right].\label{DRhoBM}
\eea
Still, if we take $b\equiv B/B'\in\mathcal{N}$, then the upper limits are related to each other as $N_2=2N_1, N_2'=2N_1'$, and $N_1'=bN_1$ in order to get a convergent result.
Eventually, by collecting the $\mu_{\rm I}$-dependent divergent parts and the spin-magnetic field coupling term, which have been subtracted in \eqref{DPiB} and \eqref{DPiBM}, and regularizing them with three-momentum cutoff $\Lambda$, we obtain a finite Ginzburg-Landau coefficient,
\bea
 \!\!\!\!\!\mathcal{A}_{\bar{\rho}_1^{+}\bar{\rho}_1^{+}} ={1\over4G_{\rm V}}\!-\!N_c\!\int^{\Lambda}\!\!\frac{d^3\mathbf{k}}{(2\pi)^3}\,
\frac{E_{\mathbf{k}}^{2}\!+\!k_{3}^{2}\!+\!m^2}{E_{\mathbf{k}}(E_{\mathbf{k}}^{2}-\mu_{\rm I}^{2}/4)}\!-\!4N_c\!\int^{\Lambda}\!\!\frac{d^4k}{(2\pi)^4}\,
\frac{eB}{k_{4}^{2}\!+\!E_{\mathbf{k}}^{2}}\,
\frac{m^2\!+\!k_{4}^{2}\!+\!k_{3}^{2}}{\left( k_{4}^{2}+E_{\mathbf{k}}^{2} \right) ^2}\!+\!\Delta\Pi_{\bar{\rho}_1^{+}\bar{\rho}_1^{+}}^{B}\!+\!\Delta\Pi_{\bar{\rho}_1^{+}\bar{\rho}_1^{+}}^{B,\mu _I}\!+\!\Pi_{\bar{\rho}_1^{+}\bar{\rho}_1^{+}}^{ B, \mu _I,T}.\label{Arho}
\eea
\end{widetext}

\section{Numerical results}\label{num}
\begin{figure}[!htb]
   \begin{center}
        \includegraphics[width=8cm]{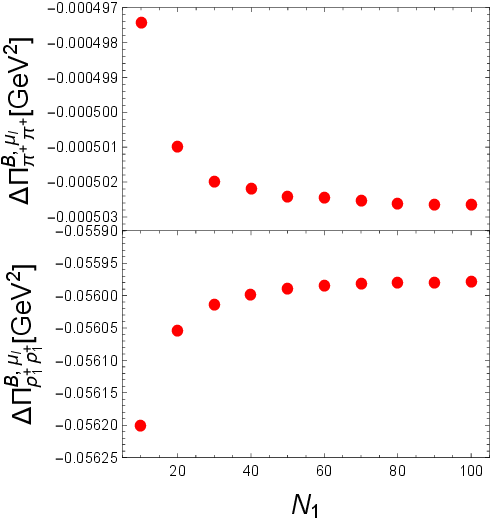}
         \caption{The $B$ and $\mu_{\rm I}$ dependent parts of the Ginzburg–Landau coefficients, $\Delta \Pi_{\pi^{+}\pi^{+}}^{B,\mu_{\rm I}}$ in \eqref{DPiBM} and $ \Delta \Pi_{\bar{\rho}_1^{+}\bar{\rho}_1^{+}}^{B,\mu_{\rm I}}$ in \eqref{DRhoBM}, as functions the upper limit $N_1$ under the regularization scheme: $N_2=2N_1, N_2'=2N_1'$, and $N_1'=bN_1$ with $b\equiv B/B'$. The physical parameters are chosen as $\mu_I = 0.6\,\mathrm{GeV}, \sqrt{eB'}=0.01\,\mathrm{GeV}$, and $\sqrt{eB}=0.2\,\mathrm{GeV}$, and the dynamical mass is solved self-consistent from the gap equation \eqref{gap} to be $m = 0.3486\,\mathrm{GeV}$. } \label{regN}
     \end{center}
\end{figure}

We now present numerical results by solving the regularized gap equation \eqref{gap} together with the Ginzburg–Landau coefficients \eqref{Api} and \eqref{Arho} for pion and rho mesons, respectively. By fitting to empirical values $m_\pi=134\,{\rm MeV}, f_\pi=93\,{\rm MeV}, \langle\bar{\psi}\psi\rangle=-2\times(250\,{\rm MeV})^3$, and a smaller $\rho$ mass, $m_\rho=600\,{\rm MeV}$, to avoid artifacts, the model parameters can be fixed as: $G_{\rm S} = 4.93\,\mathrm{GeV}^{-2}$, $G_{\rm V} = 3.37\,\mathrm{GeV}^{-2}$, $\Lambda = 0.653\,\mathrm{GeV}$, and $m_0 = 5\,\mathrm{MeV}$~\cite{Cao:2019res, Zhuang:1995uf}. To proceed, we firstly demonstrate that \eqref{DPiBM} and \eqref{DRhoBM} converge very well with increasing $N_1$ under the proposed regularization scheme, see Fig.~\ref{regN}. 

\begin{figure}[!htb]
	\begin{center}
	\includegraphics[width=8cm]{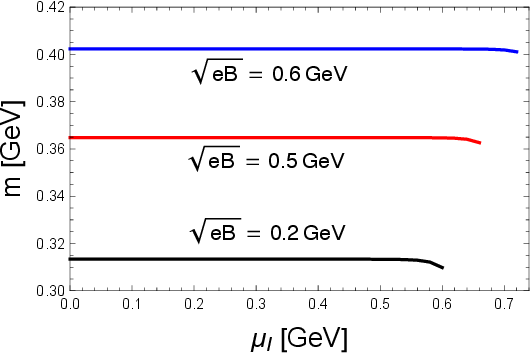}
	\caption{The dynamical quark mass $m$ as a function of isospin chemical potential $\mu_I$ for three magnetic fields, $\sqrt{eB} = 0.2$ (black), $0.5$ (red), and $0.6\,{\rm GeV}$ (blue). }\label{mass}
	\end{center}
\end{figure}

\begin{figure}[!htb]
   \begin{center}
        \includegraphics[width=8cm]{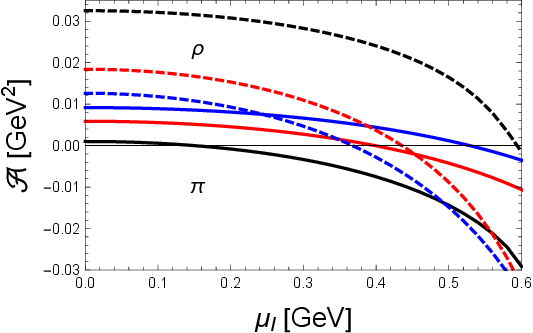}
      \caption{The Ginzburg–Landau coefficients $\mathcal{A}_{\pi^{+}\pi^{+}}$ (solid lines) and $\mathcal{A}_{\bar{\rho}_1^{+}\bar{\rho}_1^{+}}$ (dashed lines) as functions of isospin chemical potential $\mu_I$ for three magnetic fields, $\sqrt{eB}=0.2$ (black), $0.5$ (red), and $0.6\,\mathrm{GeV}$ (blue). Note that the point with $\mathcal{A}=0$ stands for the onset of instability under the assumption of a second-order transition, which then determines the phase boundary.}
    \label{GLs}
      \end{center}
\end{figure}

At zero temperature, we first solve the dynamical quark mass $m$ for three representative magnetic fields, $\sqrt{eB} = 0.2$, $0.5$, and $0.6~\mathrm{GeV}$, see Fig.~\ref{mass}. For all $B$, $m$ remains constant at low $\mu_I$ and then decreases monotonically as $\mu_I$ increases, signaling gradual restoration of chiral symmetry~\cite{He:2005nk}. And as a consequence of magnetic catalysis effect~\cite{Gusynin:1994xp}, the onset of the decreasing shifts to larger $\mu_I$ with increasing magnetic field. Usually, pion superfluidity would show up in the chiral symmetry breaking phase~\cite{Son:2000xc,Kogut:2002zg,He:2005nk}, so it is useless to extend the calculations with only chiral condensate into the chiral symmetry restoration regime here.
With respect to these magnetic fields, the Ginzburg–Landau coefficients for pion and rho mesons are shown together in Fig.~\ref{GLs}. As we can see, for a given magnetic field, the GL coefficients would always decrease from positive to negative with increasing $\mu_I$, implying the instability to pion or rho meson condensations when the isospin density is large. However, at zero $\mu_I$, the responses of pion and rho to $B$ are different: $\mathcal{A}_\pi$ increases with $B$ while $\mathcal{A}_{\bar{\rho}_1^{+}\bar{\rho}_1^{+}}$ decreases with $B$, in line with the increasing lowest energy of $\pi^+$ and decreasing lowest energy of $\bar{\rho}_1^{+}$~\cite{Cao:2020pmm}. At zero temperature, the critical $\mu_I$ is consistent with the corresponding lowest energy, so we find that the critical $\mu_I$, given by $\mathcal{A}=0$, increases with $B$ for pion superfluidity and decreases with $B$ for rho superconductivity. Eventually, the latter becomes smaller than the former around $\sqrt{eB}=0.5\,\mathrm{GeV}$, which means that rho superconductivity takes advantage over pion superfluidity when $B$ is large, consistent with our initial expectation.

\begin{figure}[!htb]
	\begin{center}
	\includegraphics[width=8cm]{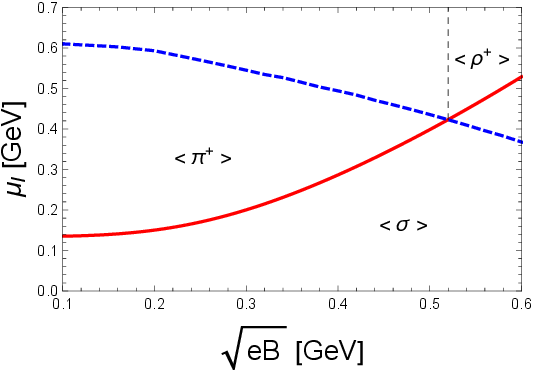}
	\caption{The phase transition lines for pion superfluidity (red solid) and rho superconductivity (blue dashed) in the $\mu_I$–$B$ plane. The notations $\langle\sigma\rangle, \langle\pi^+\rangle$, and $\langle\rho^+\rangle$ correspond to the normal chiral symmetry breaking phase, pion superfluidity, and rho superconductivity, respectively. The thin dashed line is just for schematic demonstration of the boundary between pion superfluidity and rho superconductivity.}\label{PT}
	\end{center}
\end{figure}

Finally, we present the phase transition lines for pion superfluidity and rho superconductivity in the $\mu_I$–$B$ plane, see Fig.~\ref{PT}. As there is artificial vacuum superconductivity when the magnetic field is large enough in the two-flavor NJL model, we confine ourselves to the region $\sqrt{eB}\leq 0.6\,\mathrm{GeV}$. From bottom up, we can tell that the QCD matter is in the normal chiral symmetry breaking phase when $\mu_I$ is not so large, but favors pion superfluidity for $\sqrt{eB}<0.52\,\mathrm{GeV}$ and rho superconductivity for $\sqrt{eB}>0.52\,\mathrm{GeV}$ when $\mu_I$ is large enough. Specifically, for pion superfluidity, the critical $\mu_I$ increases with $B$, consistent with the prediction of chiral perturbation theory~\cite{Adhikari:2015wva, Adhikari:2018fwm}; while for rho superconductivity, the critical $\mu_I$ decreases with $B$. Within these superconducting phases, we are not clear where is the internal boundaries or if there is a coexistent region.

\section{Summary}\label{summary}
In this work, we explore QCD phase transition at finite isospin density and magnetic field within the extended two-flavor Nambu--Jona-Lasinio model by considering both pion superfluidity and rho superconductivity. In light of the findings within chiral perturbation theory, the transitions from normal chiral breaking phase to them are expected to be of second order with increasing isospin chemical potential, so we adopt the Ginzburg-Landau approximation to study these phase transitions. Under such circumstances, the compact proper-time representation of fermion propagators would encounter artificial divergence for a large isospin chemical potential, hence we refer to the Landau representation instead to calculate the Ginzburg-Landau coefficients. To work with the Landau representation, a consistent regularization scheme is needed to deal with the divergences from the summations over Landau levels, that is, the cutoffs of Landau energies should be the same for different magnetic fields. Then, the Ginzburg-Landau coefficients for pion and rho mesons are worked out both analytically and numerically in random phase approximation. The numerical results show that pion superfluidity is favored for $\sqrt{eB}<0.52\,{\rm GeV}$ while rho superconductivity is favored for $\sqrt{eB}>0.52\,{\rm GeV}$ when increasing isospin chemical potential, in line with the magnetic enhancement (reduction) of the lowest energy of $\pi^+ ({\rho}^{+})$ meson. For rho meson with physical vacuum mass, the rho superconductivity would still occur at sufficiently large magnetic field and isospin chemical potential by following the mass inversion found in lattice QCD simulations~\cite{Bali:2017ian}, because the critical isospin chemical potential is just given by the effective meson mass at zero temperature~\cite{Son:2000xc}. However, the critical magnetic field will be even larger than that in the present case with the vacuum mass $m_\rho=600\,{\rm MeV}$.

To our knowledge, rho superconductivity has not been explored before under the circumstances with both finite magnetic field and isospin density, and its existence implies an interesting and nontrivial interplay between QCD and QED. In the future, the work can be extended to the realistic three-flavor NJL model where the artificial vacuum superconductivity can be well avoided by the splitting magnetic catalysis effect to $u$ and $d$ quarks~\cite{Cao:2019res}. Based on that, it is important but challenging to explore the transition between pion superfluidity and rho superconductivity by increasing the magnetic field at a large isospin chemical potential. As an application, these studies could shed light on possible QCD phase transitions in the early Universe~\cite{Vovchenko:2020crk,Middeldorf-Wygas:2020glx,Cao:2021gfk,Cao:2022fow,Cao:2024fyk}.

\section*{Acknowledgment}
The authors are funded by the National Natural Science Foundation of China with Grant Nos. 12447102 and 12575152, and the Natural Science Foundation of Guangdong Province with Grant No. 2024A1515011225.


\end{document}